\documentclass[runningheads,a4paper]{llncs}

\usepackage{amssymb}
\usepackage{amsmath}
\usepackage{tikz}
\usepackage{amsfonts}
\usetikzlibrary{positioning}
\usetikzlibrary{calc}
\usetikzlibrary{shapes,arrows}
\usetikzlibrary{backgrounds}
\usetikzlibrary{shapes.geometric,fit,automata}

\usepackage{pdfpages}

\usepackage{url}
\usepackage{hyperref}

\newcommand{\codesize}{\small}

\title{Context-Updates Analysis and Refinement in Chisel%
\thanks{This research has been partially supported by the
MINECO Spanish project \emph{TRACES} (TIN2015-67522-C3-3-R) and
by the Comunidad de Madrid project \emph{N-Greens Software-CM} (S2013/ICE-2731)}
\titlerunning{Context-Updates Analysis and Refinement in Chisel}
\author{Irina M\u ariuca As\u avoae \inst{1},  Mihail As\u avoae \inst{1}, 
Adri\'an Riesco\inst{2} \institute{
Inria Paris, France\\
\and
Universidad Complutense de Madrid, Spain\\
}
}
}

\toctitle{Context-Updates Analysis and Refinement in Chisel}

\begin{document}

\maketitle

\begin{abstract}
This paper presents the context-updates synthesis component of Chisel--a tool that synthesizes a program slicer directly from a given algebraic specification of a programming language operational semantics. (By context-updates we understand programming language constructs such as goto instructions or function calls.) 
The context-updates synthesis follows two directions: an overapproximating phase that extracts a set of potential context-update constructs and an underapproximating phase that refines the results of the first step by testing the behaviour of the context-updates constructs produced at the previous phase.
We use two experimental semantics that cover two types of language paradigms: high-level imperative and low-level assembly languages and we  conduct the tests on standard benchmarks used in avionics.

\smallskip

\noindent\textbf{Keywords:} generic slicing tool, programming languages formal semantics, Maude, synthesis
\end{abstract}

\section{Introduction\label{sec:intro}}

\emph{Slicing} is a well established analysis method that takes a program and a \emph{slicing criterion} (i.e., a program point \emph{pc} and a set of program variables \emph{V}) and produces a \emph{program slice} (i.e., the parts of the program containing language constructs units, usually discriminated based on the sequencing operator, that change the variables in \emph{V}, directly or indirectly,  during the program executions either up to or from the program point \emph{pc}). Note that depending on the moment slicing is applied, we have either dynamic slicing---used at the program runtime, and static slicing---used without executing the program. In this paper, we focus on static slicing and we refer it as simply \emph{slicing}. Moreover, hereafter we refer the language constructs units, i.e., the syntactic components of the programming language that are separated by sequencing operators, as \emph{instructions}.

The main idea in program slicing relies on the evaluation of the data flow equations over the control flow graph of the program. Obviously, besides the data-flow, there is a need of additional techniques to help with other language features---\cite{tip-1995-jpl} gives a comprehensive survey on the standard program slicing techniques applied over different programming language concepts such as standard imperative, pointers, unstructured control flow, and concurrency. Generally, these techniques use the programs' control flow graphs with various augmentations, e.g., the function calls are usually represented by call-edges~\cite{pnueli}.  Consequently, any programming language supporting slicing has to be automatically translated into control flow graph based models.

Field and Tip show in~\cite{field-tip-1998-ist} a method to derive program slices and dependences from \emph{term rewriting systems}. This method is applicable to any language with semantics specified as a term rewriting system. Hence, the translation of the programs into their afferent model for slicing is replaced by describing the semantics of the programming language as a rewriting system. Furthermore, the rules in this rewriting system are augmented with wrappers, which maintain the slicing information. In order to compute a program slice, the term representing the program is rewritten with the augmented rewriting system until it reaches the normal form, which contains the slice via the wrappers. This method is tantamount to determining the slice in a dynamic fashion during program execution. 

Along the lines of programming language semantics as rewriting systems, we observe an increased interest in defining various languages to cover many programming language paradigms. This desideratum is stated in \emph{the rewriting logic semantics project}~\cite{semanticsProject}, where the programming languages semantics are defined as rewriting systems using Maude,  and it is followed by the work in the $\mathbb{K}$ framework~\cite{rosu-serbanuta-2010-jlap}.
Maude~\cite{maude-book} is a high-level
language and high-performance system supporting both equational and
rewriting logic computation. Maude modules correspond to specifications
in \emph{rewriting logic}~\cite{Meseguer92-tcs}, a logic that allows specifiers
to represent many models of concurrent and distributed
systems. This logic is an extension of
\emph{membership equational logic}~\cite{BouhoulaJouannaudMeseguer00}, 
an equational logic that,  
in addition to equations, allows the statement of
\emph{membership axioms} characterizing the elements of a sort.
Rewriting logic extends membership equational logic by adding rewrite rules
that represent transitions in a concurrent system and can be nondeterministic.
In the context of~\cite{semanticsProject}, these rules correspond to the execution 
of the different instructions
in our programming language, hence allowing a natural representation for any
programming language semantics.
As a semantical framework, Maude has been used to specify the semantics of several languages,
such as LOTOS~\cite{VerdejoMartiOliet05b}, CCS~\cite{VerdejoMartiOliet05b}, and
Java~\cite{farzan-et-2004-cav}. 
Moreover, the $\mathbb{K}$-Maude compiler~\cite{rusuSemantics16},
which is able to translate $\mathbb{K}$  
specifications into Maude, has
eased the methodology to describe programming language semantics in Maude. 

Our work comes to complement the rewriting logic semantics project by developing static analysis methods, in particular slicing, for programs written in languages with an already defined rewriting logic semantics in Maude. Our approach analyses a given programming language semantics and synthesizes the necessary information for program slicing. We use the results of the syntheses to traverse the program term in order to obtain the program slice. However, we do not execute the program as in~\cite{field-tip-1998-ist}. Rather, we construct over the program an augmented control flow graph structure and we use it to obtain the program slice. The novelty, comparing to the standard methods presented in~\cite{tip-1995-jpl}, is that we construct the program models in a generic way, for any programming language with a given algebraic semantics. 

Our approach is implemented in Chisel\footnote{\url{https://github.com/ariesco/chisel}}, a Maude tool for generic program slicing~\cite{riesco-et-al-2017-fase}. Chisel takes a programming language semantics, given as a Maude specification, breaks it into pieces of interest for slicing, and uses these pieces to augment the program term and to produce the model, which is then sliced. 
For experiments we use two semantics: a semantics for an imperative programming language with functions, WhileFun, and a semantics for the MIPS assembly language. Chisel synthesizes these semantics to extract operators that produce updates at the memory level. These operators are then used to produce necessary information for slicing, e.g., side-effect instructions. The final step of Chisel is the program slicing analysis that takes a program and produces its slice w.r.t. a slicing criterion. Chisel aims to evolve into a framework for  \emph{generic static slicing}. 

The main argument for the genericity claim lays in the fact that any programming language paradigm involves a semantic notion of memory/environment which is crucial for slicing and on which we focus our syntheses. Another argument is given by Tip's survey~\cite{tip-1995-jpl} which presents specialized slicing algorithms for various programming paradigms. However, there is a price to pay for genericity: the slicing precision. Namely, the analyses of the programming language semantics produce supersets of the language constructs involved in slicing. Hence, the loss of precision directly depends on the imprecision of the synthesized language constructs. For producing more accurate synthesis results, we introduce the filtering step based on program testing.

With the current development of Chisel we target sequential imperative code without dynamic allocation that is generated from synchronous designs---a class of applications used in real-time systems, e.g., avionics. The contribution of this paper is presenting the context-updates synthesis component of Chisel, where by context-updates we understand programming language constructs such as goto instructions or function calls. 
The context-updates synthesis follows two directions: an overapproximating phase when we analyse the language semantics specification to extract a set of potential context-update constructs and an underapproximating phase when we stress-test the semantics to refine the context-updates obtained at the first step. The underapproximating  phase, firstly introduced   in this paper, is justified by the lack of precision of the overapproximating phase for the context-updates. This lack of precision is most likely due to the laxity of the automatic detection of stack-like memory operators.
Note that the class of target languages, i.e., programming languages present in the synchronous compilation chain, does not involve pointers while the arrays are always of fixed size. Since Chisel does not handle programs with pointers yet, we transform the fixed-size arrays into function calls (i.e., we add to the program a function that implements array accesses) so we can use Chisel for slicing industrial benchmarks that contain arrays.

The rest of the paper is organized as follows: in Section~\ref{sec:related} we present a comparison with existing works on generic program slicing, in Section~\ref{sec:prelim} we give an overview of the Chisel tool, in Section~\ref{sec:alg} we describe our method for context-updates synthesis and its integration in Chisel, 
and in Section~\ref{sec:exp} we describe the experimental evaluation of selected benchmarks. Section~\ref{sec:conc} concludes and outlines some future work directions.
The complete code of the tool and examples are available at
\url{https://github.com/ariesco/chisel}.

\section{Related work\label{sec:related}}

Program slicing~\cite{weiser-1981-icse} is a standard analysis technique, hence most static analyzers contain some variant of program slicing. Slicing techniques~\cite{tip-1995-jpl} are classified as static,  i.e., the slices are computed without assuming a particular program input, and dynamic, i.e., the test cases determine the program slices. 

Standard program analyzers include the necessary infrastructure to compute static slicing in the encoding of the language semantics, the control-flow graph, the dependency relations between program variables, etc. Examples of program slicing tools integrated in program analyzer are, for high-level languages: FramaC~\cite{framac} for C code, and CodeSurfer~\cite{codesurfer} and Wala~\cite{wala} for Java, whereas for low(er) level languages: Giri~\cite{giri} for LLVM intermediate representation, CodeSurfer/x86 for disassembled x86 executables~\cite{codesurferx86}, MCSLICE~\cite{srinivasan-reps-2016-oopsla} for Intel IA-32 microcode, and SlicingDroids~\cite{slicingdroids} for Android executables. All these tools translate a program into a model for analysis, then analyze the model. However, the translation phase is particular to the language for which the analyzer is built and takes into account knowledge about the particularities of each language. Chisel aims to unify the translation phase by inferring the particularities of each language from its given algebraic semantics. Through this, we explore the genericity limits of slicing, in particular, and static analysis, in general. 

An early work on generic slicing is presented in~\cite{dancic-harman-2000-sac} where the tool compiles a program into a self slicer. Generic slicing is also the focus in~\cite{binkley-et-all-2014-fse,field-ramalingan-tip-1995-popl}. The ORBS tool~\cite{binkley-et-all-2014-fse} proposes a technique for dynamic slicing, based on statement deletion. A program slice is constructed iteratively by removing statements from the original program and then checking if the transformation is semantics-preserving w.r.t.\ the slicing criterion. Their semantics preserving verification phase relies on novel testing techniques~\cite{MHarmanICSE2017}. Chisel proposes a complementary technique to the dynamic slicing of ORBS, as it computes static slicing based on in-depth investigation of the formal language semantics. We also use benchmark testing techniques for improving the precision of the context-updates synthesis.

Another generic program slicing technique is proposed in~\cite{field-ramalingan-tip-1995-popl} where an algorithm mechanically extracts slices from a common intermediate representation named PIM. The algorithm relies on a well-defined, non-trivial, and language dependent transformation between a language semantics of choice and PIM. The approach in~\cite{field-ramalingan-tip-1995-popl} is generic in the sense that notions of static and dynamic slices are represented as constrained slices and various slicing methods are collapsed in a parametric slicing procedure. Chisel integrates now only static slicing and addresses genericity from a different angle: it eliminates the need of a language-dependent translation by working directly on the language semantics.

In rewriting logic, the work in~\cite{alpuente15} implements dynamic slicing for execution traces of the Maude model checker. The semantics is executed for an initial given state, then dependency relations are computed using a backward tracing mechanism. In comparison, Chisel proposes a static approach built around a formal semantics and with an emphasis on computing slices for programs and not for given traces (e.g., of model checker runs). Also, our proposed algorithm for context-update inference is based on a notion of hypertree, which we introduced and used for side-effects analysis, in~\cite{riesco-asavoae-2012-wadt}. A similar construction to our hypertree, called 2D graph is used in~\cite{lucas-et-al-2014-lopstr} for proving termination of term rewriting systems. 

Besides the generic aspect of Chisel, we mainly address in this paper the context-updates discovery component in our framework that is also of interest in the functional programming community. Functional programming proposes richer notions of contexts (and context manipulation) than what we consider in our framework. Briefly, the standard definition of a context as variables in scopes is extended in functional languages in several directions. On one hand, there are high-level constructs such as {\it call/cc} - call with current continuation - from the Scheme language~\cite{AbelsonS85}, where snapshots of the current control states are manipulated as values (e.g., passed as arguments to function calls). One another hand, there are extended notions for contexts to capture security properties, as in the SLam calculus from~\cite{HeintzeR98} or parameters of the execution platforms~\cite{TalpinJ92,PetricekOM14}. For example, the contexts are used to track how programs affect an execution environment (e.g., the effect systems~\cite{TalpinJ92}) or the complementary approach about how programs depend on the execution environment (e.g., the coeffect systems~\cite{PetricekOM14}). In our framework, the context is a first-order variable that could be explicitly or implicitly represented in the programming languages' semantics. We identify context changes (i.e., context-updates) in a generic manner, directly from a formal language semantics given as rewrite theories. Our context-updates discovery is less particularized as the mentioned related work in functional programming. This is due to the genericity character of our approach, i.e., we do not address a particular type of memory/environment representation as the one in functional programming. Nevertheless, the rich representations of context from functional programming are of interest for our framework in order to specialize the context-updates detection with the inference of types of variables updates during context changes, e.g., different parameter passing styles at function call.

The theoretical ideas underlying Chisel are in~\cite{riesco-asavoae-2012-wadt,asavoae-riesco-2014-ifm,riesco_asavoaeLOPSTR15}. In~\cite{riesco-asavoae-2012-wadt} we describe the methodology for performing intraprocedural slicing, which is improved in~\cite{asavoae-riesco-2014-ifm} by implementing interprocedural slicing. In~\cite{riesco_asavoaeLOPSTR15} we introduce an algorithm for inferring the data-flow information to automatically detect how the language constructs work with the memory. A description of the implementation of Chisel could be found in~\cite{riesco-et-al-2017-fase}. A preliminary study of the benchmarks used for testing in this paper is presented in~\cite{asavoae-riesco-2014-ifm}, but limited to subparts of the code and only evaluated on high-level imperative languages.

\section{The Chisel system\label{sec:prelim}}
 
We briefly describe in this section the ideas underlying Chisel tool.
Chisel aims to advance the generic synthesis of program models from any programming language, provided the algebraic semantics of the language is given as a rewriting system. For now, the analysis of interest is \emph{program slicing}. Note that the standardized model used by program slicing is an augmented control flow graph, i.e., a set of control flow graphs connected by call edges. 

The crucial information used in slicing is related to the data flow: which language constructs produce the data flow and how the data is actually flowing. The main observation we use for Chisel is the fact that side-effects induce an update in the memory afferent to the program. Hence, Chisel first detects the operators used by the semantics to reproduce memory updates. Then, the usage of the memory update operators is traced through semantics up to the language constructs. Any language construct that may produce a memory update is classified as producing side-effects. Moreover, following the direction of the memory updates, we infer also the data flow details (i.e., source-destination) for each side-effect language construct. Finally, the information gathered by Chisel about language constructs is used to traverse the term representing the program and to extract the subterms representing the slice.

\tikzstyle{block} = [rectangle, draw, fill=blue!10, 
    			     text width=7.5em, text centered, 
			     rounded corners, minimum height=3em]
\tikzstyle{comm} = [rectangle, draw, fill=white, 
    			     text width=9em, text centered, dashed,
			     rounded corners, minimum height=3em]				     
\tikzstyle{line} = [draw, -latex']
\tikzstyle{cloud} = [draw, ellipse,fill=red!10, node distance=4cm, 
			     text width=7em, text centered,
			     minimum height=3em]

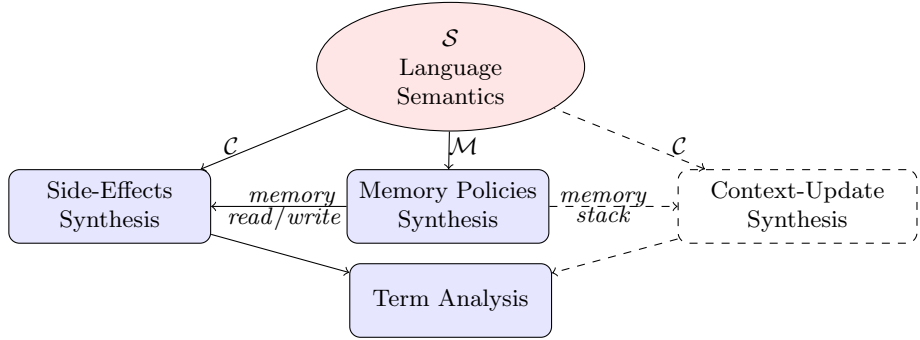
\begin{figure*}[!t]
\begin{center}
	\begin{tikzpicture}[node distance = 2cm, auto]
    	\node [block] (sideeffect) {Side-Effects Synthesis};
	\node [block, right of=sideeffect, node distance=4.45cm] 
			(mempol) {Memory Policies Synthesis};
	\node [comm, right of=mempol, node distance=4.6cm] 
			(contextupdate) {Context-Update Synthesis};
    	\node [cloud, above right=0.75cm and 1.9cm of sideeffect] 
			(model) {$\mathcal{S}$\\ Language Semantics};
    	\node [block, below of=model, node distance=3.1cm] 
	                 (termanalysis) {Term Analysis};
    	\draw [<-] (mempol) -- (model);
	\draw [<-] (sideeffect) -- (model);
	\draw [<-,dashed] (contextupdate) -- (model);
	\draw [->] (sideeffect) -- (termanalysis);
	\draw [->, dashed] (contextupdate) -- (termanalysis);
    	\draw [->] (mempol.west) -- (sideeffect.east);
    	\draw [->, dashed] (mempol) -- (contextupdate);

	\draw (1.6,0.8) node {$\mathcal{C}$};
	\draw (4.65,0.8) node {$\mathcal{M}$};
	\draw (7.5,0.8) node {$\mathcal{C}$};
	\draw (2.4,0.1) node {${\it memory}$};
        \draw (2.3,-0.15) node {${\it read/write}$};
        	\draw (6.5,0.1) node {${\it memory}$};
        \draw (6.5,-0.1) node {${\it stack}$};
	\end{tikzpicture}
\end{center}
\vspace{-0.5cm}
\caption{\label{f:components} Chisel components: the formal language semantics and the analyses.}
\vspace{-0.5cm}
\end{figure*}

The tool works under a few assumptions w.r.t.\ $\mathcal{S}$--\emph{the programming language semantics specification}. Firstly, we assume that $\mathcal{S}$ is provided as an algebraic specification in rewriting logic. Given the bundle of work in the area of programming language specifications using rewriting logic, as discussed in Section~\ref{sec:intro}, we consider that this first assumption does not impose a restriction on the generality. Secondly, we assume the existence of a certain structure in this semantics. Namely, the instructions are terms of a particular sort (i.e.,\ a type) and the memory/environment/machine on which the programs are running are described by the operators defined in a certain specification module. The idea behind this assumption is the fact that any semantics of a programming language uses an (abstract) memory and some operations over this memory. 

We present in Fig.~\ref{f:components} the structure of Chisel: its components and their input-output relations. We briefly address each of the components in the following subsections, except the context-update synthesis, which is the main contribution of the current work and is to be elaborated in the remaining  of this paper.

\subsection{Memory policies synthesis\label{subsec:mem-pol}}
Let us denote $\mathcal{M}$ as the part of $\mathcal{S}$ that defines some (abstract) form of the memory used during program execution. 
Our assumption about the structure of the memory is that it connects the variables in the program with their values possibly via a chain of intermediate addresses. We define a \emph{memory policy} as a particular type of operators specified using $\mathcal{M}=\{o:w\rightarrow s\}$, where $w$ and $s$ are standardly denoted as the arity and, respectively, co-arity of $o$. (Note that $s$ denotes a sort while $w$ denotes a list of sorts.) For example, a \emph{memory-read} is the set of operators in $\mathcal{M}$ that contain in their arity the sort for variables and for memory and in their co-arity the sort for values. A \emph{memory-write} operator contains in its arity sorts for memory, variables, and values, and in its co-arity the memory sort. Also, the rules defining this operator change the memory by updating the variable with the value. In~\cite{riesco_asavoaeLOPSTR15} we present the analyses of $\mathcal{S}$ for read/write memory policies that produce the set of operators that manipulate the memory according to the given policy.

\subsection{Side-effect synthesis\label{subsec:th_se_sem}}
Let us denote by $\mathcal{C}$ the part of $\mathcal{S}$ that defines the operators representing the programming language constructs, i.e., language instructions. The first inference that Chisel has to make is regarding which elements in $\mathcal{C}$ modify the program variables. Since in the memory $\mathcal{M}$ the program variables are connected to their values, determining constructs that modify a variable essentially means tracking the effects of an instruction over its variable component until reaching the memory level. Hence, we name \emph{side-effect} language constructs those operators in $\mathcal{C}$ that produce a memory-write over some of its variable component.

To determine the subset of $\mathcal{C}$ that may be side-effect constructs, Chisel identifies the set of rules $\mathcal{R}$ in $\mathcal{S}$ containing in the left-hand side (as a subterm) the $\mathcal{C}$ operators. The side-effect synthesis starts with the rules in $\mathcal{R}$ and constructs a \emph{hyper-tree} $\mathcal{T}$ whose nodes are sets of rewrite rules and edges are unification based dependencies between these rules. Chisel discriminates the side-effect constructs by following the paths in this hyper-tree from the root to the leaves. The paths $\mathcal{P}$ leading to leaves that contain rules already classified by the memory policy phase as memory-writes are signalling the side-effect constructs. This part of Chisel is presented in~\cite{riesco-asavoae-2012-wadt}. 

The next phase of the side-effect synthesis consists in using the constructed $\mathcal{T}$ to determine the data flow (source-destination) produced by the side-effect constructs. Essentially, at this phase, Chisel trickles-up the paths $\mathcal{P}$ of the hyper-tree $\mathcal{T}$, starting from the leaves up to the root. Namely, at the leaves level we identify the variable subterm as destination and the value as the source. This identification is based on the read/write memory policy phase. The information regarding the source-destination relation between these two subterms (i.e., value/variable) is propagated up on each path in $\mathcal{P}$ by a backwards inference of the unifications of these subterms. When reaching the root of $\mathcal{T}$, the value at the memory level is hooked 
to the sources subterms and the variable to the destination subterms. Hence, we determine the data flow induced by each side-effect construct and we describe this as a part of~\cite{riesco_asavoaeLOPSTR15}.

Note that side-effects synthesis determines an over-approximation of the side-effect constructs. The data flow inference phase only enriches each of the already discovered side-effects constructs with key information for the program slicing, i.e., the source-destination direction of the flow of data.

\subsection{Term analysis\label{subsec:term-travers}}
The algorithm for slicing a program $p$ takes as input a \emph{slicing criterion} $S$ consisting in a set of program variables. In this step, Chisel takes the tree $T_p$  representing the program term and traverses it repeatedly. Each traversal phase adds new elements to the set $S$ and the process is repeated until the set $S$ stabilizes. While traversing $T_p$, whenever a side-effect construct is encountered, if the destination of this construct is from $S$ then all the source variables are added to $S$. Also, whenever a context-update construct is encountered, the traversal of $T_p$ is redirected towards the $T_p$'s subtree whose root matches a particular subterm of the context-update construct. At the end of the traversal, the program slice is given by the skeleton of $T_p$ containing the subtrees representing the instructions that produced changes to the set $S$.

\section{The context-update inference algorithm\label{sec:alg}}

In this section we present our approach towards discovering context-update constructs in the programming language under consideration. We start by setting some notation and defining the intuitive ideas. Note that in the followings we use notation introduced in the previous section.

Firstly, we denote $L_p$ the list of elements from $\mathcal{C}$--the language instructions' sort--obtained by a preorder traversal of $T_p$--the tree associated to the program $p$. We define as \emph{context-update constructs} (context-updates for short) those operators in $\mathcal{C}$ that, during program execution using $\mathcal{S}$, produce changes to the list $L_p$. For example, function calls and gotos are context-update constructs. We denote by \emph{context-updates synthesis} the strategy of deducing, based on the language semantics $\mathcal{S}$, an overapproximation of the set of context-update instructions. 
 
The methodology we propose for context-updates synthesis follows the same strategy as the one for side-effects described in the previous Sections~\ref{subsec:mem-pol} and~\ref{subsec:th_se_sem}. Namely, we firstly apply sort-based patterns to the memory module in $\mathcal{S}$ in order to identify stack structures/memory operators or, short, \emph{memory-stacks}  (Section~\ref{smp}). Secondly, using the memory-stacks we traverse the hyper-tree $\mathcal{T}$ to discover the set $\mathcal{O}$ of language constructs that directly use the memory-stacks (Section~\ref{cui}). Note that $\mathcal{O}$ is an overapproximation of the context-updates since our target semantics $\mathcal{S}$ describe context-free languages of either high or low level. As the context-free languages need some stack representation and we trigger our context-updates synthesis by an initial phase that discovers memory-stack patterns in $\mathcal{S}$. Finally, in order to make the set $\mathcal{O}$ more accurate we use a refinement step that, based on the execution of benchmarks, partitions the subset $\mathcal{O}$ in three: $\mathcal{O}_f$ the function call constructs, $\mathcal{O}_g$ the goto constructs, and $\mathcal{O}_r$ the residue constructs that are present in $\mathcal{O}$ due to the overapproximations in the first two steps (Section~\ref{cur}).
 
\subsection{Memory-stack policy\label{smp}}

The memory-stack policy determines rules in $\mathcal{S}$ constructing stack-like structures at the memory level. The strategy applied for memory-stack policy is similar to the strategy described in Section~\ref{subsec:mem-pol}. Namely, we have two patterns we search for: explicit and implicit. The explicit memory-stack policy where we determine non-commutative memory operators $s: S S' {\rightarrow} S$ or $s: S' S {\rightarrow} S$ that have a subsorted arity $S'\leq S$ and all the rules describing them either add or substract one element. The implicit pattern uses the conditional rules over the language semantics to produce memory-stacks. The implicit pattern is produced by the Maude's evaluation semantics that uses a an evaluation stack for conditional rules. Namely, the evaluation of the conditional rule's body (i.e., the statement between the \texttt{crl} and \texttt{if} keywords) is postponed until the evaluation of the rule's condition (i.e., the statement after \texttt{if} keyword) is completed.

\begin{example}\label{ex:mem-mod}
We present in this example the memory specification for WhileFun--an imperative
language with assignment, conditional, loops, local variables,
an input/output buffer, and function calls~\cite{Hennessy90,asavoae-riesco-2014-ifm}. 
Assuming we have defined the syntax for the
language in a module \texttt{WHILE-SYNTAX} (which includes definitions for variables,
Boolean values, and numeric values), the module \texttt{MEMORY} imports this module and
defines the sorts \texttt{Env} for the environment, which maps variables
to values, and \texttt{ESt} for a stack of environment, which will be used when a new
context is required:
{\codesize
\begin{verbatim}
fmod MEMORY is
  pr WHILE-SYNTAX .
  sorts Env ESt .
  subsort Env < ESt .
  ...
\end{verbatim}
}
\noindent
where the \texttt{subsort} indicates that a single environment states for a singleton stack, i.e., the environment type is a subtype of the environments' stack .
Constructors of these sorts are defined by using \verb"op" and the attribute \verb"ctor".
In this case, we define the empty environment (\texttt{mt}); a single assignment,
which receives a variable and a value (underscores are placeholders); and the composition
of environment, defined with empty syntax and defined as commutative and associative
and having \texttt{mt} as identity:
{\codesize
\begin{verbatim}
  op mt : -> Env [ctor] .
  op _=_ : Variable Value -> Env [ctor] .
  op __ : Env Env -> Env [ctor comm assoc id: mt] .
\end{verbatim}
}

\noindent
Similarly, the stack is built by putting together stacks with the \verb"_|_" operator:
{\codesize
\begin{verbatim}
  op _|_ : ESt ESt -> ESt [ctor assoc] .
\end{verbatim}
}
\noindent
The operator \texttt{\_|\_} follows the explicit memory-stack policy and it will be used in the context-update synthesis, as described in Example~\ref{ex:hyper-tree} from Section~\ref{cui}. The memory module also contains functions for variables' update, variables' look-up, and new variables allocation. Below we show Chisel commands and the results for the memory-stack policy applied on WhileFun:
{\codesize
\begin{verbatim}
Maude> (memory inferences .)
ESt RWBUF
Maude> (context update sorts .)
ESt
Maude> (memory-stack ops .)
_|_
\end{verbatim}
}
\noindent
Namely, \texttt{memory inferences} command produces the sorts that agree with memory-stack policy: \texttt{ESt}--the environment stack sort defined in the \texttt{MEMORY} module--and \texttt{RWBUF}--the sort defining the memory buffer that handles the results of the read/write instructions in WhileFun. The \texttt{context update sorts} command keeps from the memory-stack sorts only the sorts that produce context changes, as we describe next, in Section~\ref{cui}. 
\end{example}

\subsection{Context-updates synthesis\label{cui}}

The synthesis of a set $\mathcal{O}$ of constructs that may produce context-updates relies on the hyper-tree constructed for the operators in $\mathcal{C}$ and is similar with the side-effects synthesis described in Section~\ref{subsec:th_se_sem} and~\cite{riesco-asavoae-2012-wadt}. The difference here is the fact that at the leaves level we now use a different memory policy (the memory-stack policy)  to filter the paths leading to context-updates. The algorithm implementing this in Chisel is defined by the operator \texttt{traverseHypertree} given in Fig.~\ref{f:traverseHypertree}:

\begin{figure}[ht]  
\begin{center}
{\scriptsize
\begin{verbatim}
  op traverseHypertree : Module QidSet TermList ContextUpdates HypertreeTraversalResult
                         -> HypertreeTraversalResult .                       
  eq traverseHypertree(M, none, TL, CU, HTR) = HTR .
  ceq traverseHypertree(M, Q ; QS, TL, CU, HTR) = traverseHypertree(M, QS, TL, CU, HTR')
   if Q in CU /\
      HTR' := add2orange(Q, HTR) .
  ceq traverseHypertree(M, Q ; QS, TL, CU, HTR) = traverseHypertree(M, QS, TL, CU, HTR)
   if traversed?(Q, HTR) .
  ceq traverseHypertree(M, Q ; QS, TL, CU, HTR) =
            if allOrange?(HTR') and not emptyHypernode(M, COND, (T, TL))
            then add2orange(Q, HTR')
            else add2olive(Q, HTR')
            fi
   if COND := getCondition(M,Q) /\ not Q in CU /\ not traversed?(Q,HTR) /\ T := getLHS(M,Q) /\
      HTR' := traverseCond(M, COND, (T, TL), CU, setAllOrangeVar(true, HTR)) .

  op traverseCond : Module Condition TermList ContextUpdates HypertreeTraversalResult
                    -> HypertreeTraversalResult .
  eq traverseCond(M, nil, TL, CU, HTR) = setAllOrangeVar(false, HTR) .
  eq traverseCond(M, T = T' /\ COND, TL, CU, HTR) = traverseCond(M, COND, TL, CU, HTR) .
  eq traverseCond(M, T := T' /\ COND, TL, CU, HTR) = traverseCond(M, COND, TL, CU, HTR) .
  eq traverseCond(M, T : S /\ COND, TL, CU, HTR) = traverseCond(M, COND, TL, CU, HTR) .
  ceq traverseCond(M, T => T' /\ COND, TL, CU, HTR) = combineHypernodes(HTR', HTR'')
   if TV := freshTerm(T) /\  
      QS := getRulesUnifying(M, TV, getRls(M), TL) /\
      HTR' := traverseHypertree(M, QS, TL, CU, HTR) /\
      HTR'' := traverseCond(M, COND, TL, CU, setAllOrangeVar(true, HTR')) .
\end{verbatim}
}
\end{center}
\caption{\label{f:traverseHypertree} The traverseHypertree operator in Chisel.}
\end{figure}

The operator in Fig.~\ref{f:traverseHypertree} computes the set of basic syntactic language constructs that {\it may} be context-updates, by inspecting the conditions and the right-hand side of each rewrite rule in $\mathcal{C}$ represented here as \texttt{Q}, i.e., the rule label. The operator unfolds the rewrite rules into the hyper-tree $\mathcal{T}$ with children nodes representing lists of rules that unify with subterms of \texttt{Q}'s conditions (each subterm of \texttt{Q} unifies with a particular node). 
The \texttt{traversalHypertree} operator goes horizontally in $\mathcal{T}$ if there is no subtree rooted in the current \texttt{Q} node. Otherwise, when $\texttt{Q}$ is the root of a subtree in $\mathcal{T}$ (e.g., when the rule \texttt{Q} is conditional), the traversal goes vertically via the operator  \texttt{traverseCond}.
The  \texttt{traversalHypertree} operator assigns each rule label \texttt{Q} to a particular set, either \texttt{orange} or \texttt{olive}, where these sets are defined as follows:
$$\begin{array}{l}
{\tt orangeSet} := \{{\tt Q}\in {\it nodes(\mathcal{T})}\ |\ \exists {\tt Q'}\in {\it subtree}({\tt Q}, \mathcal{T}): {\tt Q'}\in {\tt ContextUpdates}\}\\
{\tt oliveSet} := \{{\tt Q}\in {\it nodes(\mathcal{T})}\ |\ \forall {\tt Q'}\in {\it subtree}({\tt Q}, \mathcal{T}): {\tt Q'}\notin {\tt ContextUpdates}\}
\end{array}
$$
\noindent
The \texttt{orangeSet} contains \texttt{Q}s that are the root of a subtree containing context-updates while \texttt{oliveSet} is context-updates free. Note that the termination of the algorithm in Fig.~\ref{f:traverseHypertree} is ensured by the fact that the specification $\mathcal{S}$ has a finite number of rules, and that any rule in $\mathcal{T}$ that was already added to either \texttt{orange} or \texttt{olive} set is not unfolded anymore. 
We give next an example that provides the intuition about the synthesis process.

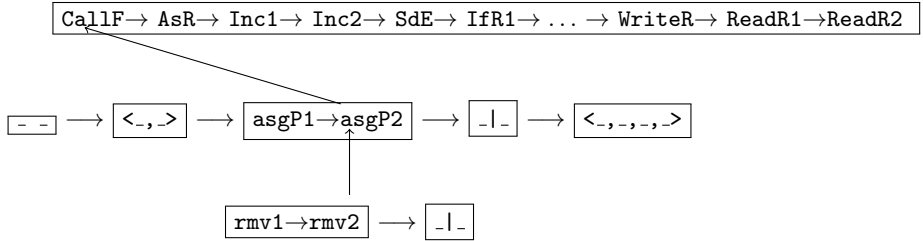
\begin{figure}[ht]  
\begin{center}
\begin{tikzpicture}
  [scale=.9,auto=left]
  \node (n5) at (7,5)  {\fbox{{\small 
                       \texttt{CallF}$\rightarrow$
                       \texttt{AsR}$\rightarrow$
                       \texttt{Inc1}$\rightarrow$
                       \texttt{Inc2}$\rightarrow$
                       \texttt{SdE}$\rightarrow $  
                       \texttt{IfR1}$\rightarrow \ldots$
                       $\rightarrow$
                       \texttt{WriteR}$\rightarrow$
                       \texttt{ReadR1}$\rightarrow$\texttt{ReadR2}} }};
  \node(n5AsR) at (1,4.9) {};
  \node (n4) at (5,3.5)  
  {
    \fbox{{\small\texttt{\_ \_}}}
   $\longrightarrow$
   \fbox{{\small\texttt{<\_,\_>}}}
   $\longrightarrow$
   \fbox{{\small \texttt{asgP1}$\rightarrow$\texttt{asgP2}}}
   $\longrightarrow$
   \fbox{{\small \texttt{\_|\_}}}
   $\longrightarrow$
   \fbox{{\small \texttt{<\_,\_,\_,\_>}}}

  };
  \node(n4VarR) at (3.6,3.5) {};
  \node(n4OpR) at (5,3.5) {};
  \node(n4upd) at (5,3.7) {};

 \node(n1Miss) at (5,2) {\fbox{\small \texttt{rmv1}$\rightarrow$\texttt{rmv2}}
                                      $\longrightarrow$ \fbox{{\small \texttt{\_|\_}}}};
 
  \foreach \from/\to in {n4upd/n5AsR, n1Miss/n4OpR}
    \draw [->] (\from) -- (\to);

\end{tikzpicture}
\end{center}
\caption{\label{f:hyper-tree} The hyper-tree constructed for WhileFun.}
\end{figure}

\begin{example}\label{ex:hyper-tree}

The first part of the hyper-tree $\mathcal{T}_{\rm WhileFun}$, constructed for WhileFun semantics, is depicted in Fig.~\ref{f:hyper-tree}. The memory-stack operator discovered here at the leaves level is \texttt{\_|\_} that is obtained by the explicit memory-stack policy. The root of $\mathcal{T}_{\rm WhileFun}$  contains the language constructs $\mathcal{C}$ where we show first \texttt{CallF} the rule label that specifies the semantics of a function call such as: 

{\codesize
\begin{verbatim}
  crl [CallF] : 
   < Call fn(actPrms), st, rwb, fs > => < skip, st'', rwb', fs >
   if fn(Prms){ C } fs' := fs /\
      < actPrms, st > => vals /\
      st' := assignPrms(actPrms, Prms, st | mt) /\
      < C, st', rwb, fs > => < skip, st'' | lenv', rwb', fs >  .
\end{verbatim}
}

The first condition in the rule \texttt{CallF} extracts
the function definition from the function set \texttt{fs} by means of a matching condition;
the second condition evaluates the arguments passed to the function;
the third condition uses the function \texttt{assignPrms} (described below) to bind the parameters to the values previously obtained; and the fourth condition evaluates the body of the function
in the new stack of environments.

{\codesize
\begin{verbatim}
  op assignPrms : ExpL VarL ESt -> ESt .
  eq [asgP1]  : assignPrms(nv, nv, ro) = ro .
  eq [asgP2]  : assignPrms((N,EL), (X,VVs), mu | ro) =
                assignPrms(EL, VVs, mu | remove(ro, X) (X = N)) .
\end{verbatim}
}

The function \texttt{assignPrms} is in charge of assigning the appropriate values
to the parameters of a function call. It receives a list of expressions,
a list of variables, and a stack of environments as
arguments and traverses the lists removing the previous value associated to the variable
at the top of the stack and binding it to the new one.

\end{example}

\subsection{Context-updates refinement\label{cur}}

For the refinement step we use a modified version of the Maude testing tool presented in~\cite{wadt10}, which  generates test cases for Maude functional modules and executes these tests while checking the conformance of their result w.r.t. a given specification $\varphi$. In our case, $\varphi$ is defined over the execution trace as defined next. 

Given the programming language semantics $\mathcal{S}$ and a program $p$, which is associated in $\mathcal{S}$ with a term with a tree representation $T_p$, we define $L_p$ the flattening of $p$ into a list of instructions (i.e., unit elements in $\mathcal{C}$) obtained by the preorder traversal of $T_p$ (i.e., the listing of the program's code instructions). 
Given a set of execution traces $E$ we denote its elements by $\varpi$, i.e., an execution path of $p$ w.r.t. $\mathcal{S}$. Furthermore, we denote by $\pi$ the filtering of $\varpi$ w.r.t. the language constructs $\mathcal{C}$. We use the standard notation for $\pi$, namely $|\pi|$ represents the length of the path, while $\pi_i, i \in \{0, \ldots, |\pi|\}$, represents the $i$-th element of the path. Note that $\pi_0$ is $\epsilon$, the empty execution list. We also denote by 
$[L_p]_{\it fn}$ the set function definitions in $p$:
$$
\begin{array}{ll}
\{L_p(k)..L_p(k+n-1)\ |& L_p(k) \in \mathcal{C}_{\it fn}\ {\rm and}\ (L_p(k+n) \in \mathcal{C}_{\it fn}\ {\rm or}\ L_p(k+n)=\epsilon)\\
& \ {\rm and}\ \forall i=k+1..k+n-1: L_p(i)\notin \mathcal{C}_{\it fn}\}
\end{array}
$$
\noindent
where $L_p(i)$ represents the $i$-th element of the list $L_p$ and $\mathcal{C}_{\it fn}$ is the set of program constructs representing function declarations.

\begin{definition}
The property $\varphi$ w.r.t. $E$ is defined as follows:\\[0.1pt]

\begin{math}
\begin{array}{l}
\forall \varsigma \in \mathcal{O},\
\forall \varpi\in E, \pi:={\it filter}_{\mathcal{C}}(\varpi),\ \forall i\in 1..|\pi |: \pi_i=\varsigma \implies\\
\ \ \ ( \pi_{i-1}\pi_i \in L_p \implies \varsigma\in \mathcal{O}_r)\land\\
\ \ \ ( \pi_{i-1}\pi_i \notin L_p \land (\pi_{i-1}, \pi_i)\in [L_p]_{\it fn})\implies \varsigma\in \mathcal{O}_g )\land\\
\ \ \ ( \pi_{i-1}\pi_i \notin L_p \land (\pi_{i-1}, \pi_i)\notin [L_p]_{\it fn})\implies \varsigma\in \mathcal{O}_f )
\end{array}
\end{math}

\end{definition}

Hence the three sets $\mathcal{O}_f$ (the function call constructs), $\mathcal{O}_g$ (the goto constructs),
and $\mathcal{O}_r$ (the residue constructs) are obtained from $\mathcal{O}$ by a discrimination process based on the analysis of testing traces. Namely, 
the residues $\mathcal{O}_r$ are constructs that execute always in programs' sequential order; the gotos $\mathcal{O}_g$ and function calls $\mathcal{O}_f$ are constructs that break the sequential order for either jumping inside the current function body, or to another function, respectively. Note that if the sets $\mathcal{O}_r$, $\mathcal{O}_g$, and $\mathcal{O}_f$ do not form a partition we use the remaining elements in $\mathcal{O}$ to signal counterexamples for the context-updates inference phase. In the next section we describe the benchmark tests we used for the experimental semantics WhileFun and MIPS.

\section{Experiments in Chisel\label{sec:exp}}

\begin{figure}[!t]

\begin{tabular}{l  l}
\begin{minipage}{6cm}
{\renewcommand{\arraystretch}{1.3}
\begin{tabular}{|l|}
\hline
{\bf Tasks : {\tt fbw} and {\tt autopilot}}\\
\hline
{\bf T1}\hspace{0.18cm} - { receive\_radio\_commands}\\
\hline
{\bf T2}\hspace{0.18cm} - { send\_data\_to\_autopilot}\\
\hline
{\bf T3}\hspace{0.18cm} - { receive\_data\_from\_autopilot}\\
\hline
{\bf T4}\hspace{0.18cm} - { transmit\_servos}\\
\hline
{\bf T5}\hspace{0.18cm} - { check\_failsafe}\\
\hline
{\bf T6}\hspace{0.18cm} - { manage\_radio\_commands}\\
\hline
{\bf T7}\hspace{0.18cm} - { control\_stabilization}\\
\hline
{\bf T8}\hspace{0.18cm} - { send\_data\_to\_fbw}\\
\hline
{\bf T9}\hspace{0.18cm} - { receive\_gps\_data}\\
\hline
{\bf T10} - { control\_navigation}\\
\hline
{\bf T11} - { control\_altitude}\\
\hline
{\bf T12} - { control\_climb}\\
\hline
{\bf T13} - { manage\_reporting}\\
\hline
\end{tabular}}
\end{minipage}
&
\begin{minipage}{2.8cm}
\includegraphics[width=4.5cm]{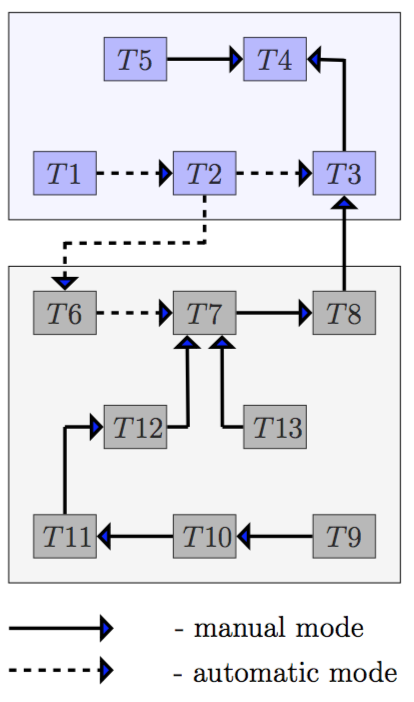}
\end{minipage}		
\end{tabular}%
\caption{\label{t:tasks} PapaBench: The tasks and their dependencies}
\end{figure}

We apply Chisel, extended with the synthesis algorithm for context-updates  on a standard benchmark for real-time systems called PapaBench~\cite{nemer-et-2006-WCET}. We consider, in our experimental evaluation, both formal semantics of WhileFun and MIPS. PapaBench is a code snapshot extracted from an actual real-time designed for Unmanned Aerial Vehicle (UAV). It consists of two communicating applications, a command management called {\tt fly\_by\_wire} and a navigation management called {\tt autopilot}. Both applications have a number of inter-dependent tasks which are executed in a control loop, at different frequencies. Structurally, {\tt fly\_by\_wire} has five tasks, named from T1 to T5 and {\tt autopilot} has eight tasks, from T6 to T13. Moreover, each application serves three interrupts, which are not of concern for the current tool evaluation. The tasks are summarized in Fig.~\ref{t:tasks} (left) and their inter-dependencies shown in the same figure (right). PapaBench application has two modes--manual and automatic. In the manual mode, the radio command (T1) of {\tt fly\_by\_wire} is executed, sending data (T2) to {\tt autopilot} which analyzes and sends back information (T6, T7, T8) to {\tt fly\_by\_wire} for processing and issuing commands (T3, T4). The automatic mode is triggered in {\tt autopilot} by the GPS communication (T9) and enables navigation, altitude and climb control (T10, T11, T12) before stabilization (T7). T5 of {\tt fly\_by\_wire} and T13 of {\tt autopilot} handle failure checking and respectively parameter reporting. 

\begin{figure*}[!t]
\begin{center}
{\renewcommand{\arraystretch}{1}
\begin{tabular}{|c|c|c|c|c|c|c|}
\hline
{\bf  Name}&{\bf \# Funs}&{\bf \# Calls}&{\bf LOC}&{\bf red (\%)}&{\bf LOC}&{\bf{red (\%)}}\\
& & &\ {\bf (WhileFun)}&\ {\bf (WhileFun)}&\ {\bf (MIPS)}&\ {\bf (MIPS)}\\
\hline
${\tt scheduler\_fbw}\ $&{14}&{18}&{103}&{72.8 \%}&{396}&{44.4 \%}\\
\hline
${\tt periodic\_auto}\ $&{21}&{80}&{225}&{73.3 \%}&{779}&{36.3 \%}\\
\hline
\hline
${\tt\bf fly\_by\_wire}\ $&{\bf 41}&{\bf 110}&{\bf 638}&{\bf 91.1 \%}&{\bf 1913}&{\bf 41 \%}\\
\hline
\hline
${\tt T1}$&{10}&{26}&{119}&{76.5 \%}&{534}&{36.2 \%}\\ 
\hline
${\tt T2}$&{9}&{9}&{59}&{69.5 \%}&{329}&{44.4 \%}\\ 
\hline
${\tt T3}$&{9}&{24}&{82}&{76.5 \%}&{501}&{43.6 \%}\\ 
\hline
${\tt T4}$&{9}&{14}&{50}&{61.5 \%}&{235}&{34.5 \%}\\ 
\hline
${\tt T5}$&{7}&{22}&{66}&{67 \%}&{453}&{51 \%}\\
\hline
\hline
${\tt\bf autopilot}\ $&{\bf 95}&{\bf 214}&{\bf 1384}&{\bf 92 \%}&{\bf 5639}&{\bf 41.5 \%}\\
\hline
\hline
${\tt T6}$&{36}&{71}&{306}&{77.2 \%}&{1329}&{54 \%}\\
\hline
${\tt T7}$&{9}&{13}&{57}&{70 \%}&{426}&{42 \%}\\
\hline
${\tt T8}$&{7}&{15}&{54}&{69.2 \%}&{219}&{38 \%}\\
\hline
${\tt T9}$&{15}&{30}&{87}&{75 \%}&{617}&{36.5 \%}\\
\hline
${\tt T10}$&{18}&{27}&{102}&{71.1 \%}&{1002}&{42.2 \%}\\
\hline
${\tt T11}$&{3}&{2}&{15}&{63.4 \%}&{90}&{70.6 \%}\\
\hline
${\tt T12}$&{4}&{3}&{49}&{66.2 \%}&{363}&{50 \%}\\
\hline
${\tt T13}$&{37}&{93}&{240}&{79.7 \%}&{1535}&{42 \%}\\
\hline
\end{tabular}
}
\end{center}
\caption{\label{f:bench} Chisel performance on PapaBench benchmark}
\end{figure*}

PapaBench is organized in modes and tasks, coordinated by a set of global variables. We apply Chisel on two programming levels: the original imperative code and the binary code obtained after disassembling. In this way we aim to study the analyzability and traceability properties of PapaBench, e.g., isolate and quantify different functionalities within each task, as well as the inter-task behavior between communicating tasks. We report the results of Chisel as reduction in the number of instructions (LOC). Next, we elaborate on experimentation (i.e., platform, organization, test cases) and its results, as summarized in Fig.~\ref{f:bench}. 

We conduct our experiments on the following settings: we run Chisel with Maude (and Full-Maude) 2.7 on a MacBook Pro 2.5 GHz, 4GB RAM, with PapaBench version 0.4 (for the WhileFun code) and the {\tt gcc 4.7.1} cross-compiler to obtain MIPS code (and with sufficient traceability to check the corresponding program slices at the high- and low-levels). 

We organize the benchmark as follows, in Fig.~\ref{f:bench}: each of the 13 tasks (the rows T1 to T13), the core functionalities of {\tt fly\_by\_wire} and {\tt autopilot} (the rows {\tt scheduler\_fbw} and {\tt periodic\_auto}), and the complete PapaBench benchmark (the rows {\tt fly\_by\_wire} and {\tt autopilot}). These latter four functionalities were introduced in~\cite{riesco-et-al-2017-fase} and included here for completion purposes. Note that in~\cite{riesco-et-al-2017-fase} the context-updates were manually introduced for each language while here we detect them automatically. The context-updates synthesis phase produces exact results for WhileFun while for MIPS the overapproximation at the synthesis phase is too large (the synthesized context-updates for MIPS include most of the language instructions). Hence, the testing phase, which is underapproximating the synthesized context-updates, is essential for context-updates in MIPS. We employ random testing on the currently described benchmarks and we reduce the context-updates for MIPS to the exact set. Hence, the results reported in the Fig.~\ref{f:bench} coincide with the results obtained in~\cite{riesco-et-al-2017-fase} where the context-updates were manually provided. We quantify the number of functions and function calls (columns {\tt \#Funs} and respectively {\tt \#Calls}), the code size ({\tt LOC}) and the slicing reduction factor, {\tt red(\%)} for both WhileFun and MIPS programs. 

The reduction factor captures the slicing performance w.r.t., the original code on both WhileFun and MIPS variants. We measure the slicing performance in the following way:
\begin{itemize}
\item[-] The rows with the full benchmark ({\tt autopilot} and {\tt fly\_by\_wire}) and its core functionalities ({\tt scheduler\_fbw} and {\tt periodic\_auto}): the slicing procedure considers as slicing criteria sets of global variables used to activate modes and inter-task communication. The {\tt red(\%)} shows the reduction percentage resulting from program slices size over the reference (original) code size.  
\item[-] The rows corresponding to each task T1 to T13 (with the exception of T11--control\_altitude--which is very small, but included for completeness purposes): the slicing procedure is based on 7 slicing criteria designed to measure several aspects of a task code. These criteria correspond to: 
\begin{itemize}
\item[1-] {\it global functionality}, i.e., variable(s) responsible for the task functionality; 
\item[2,3-] {\it mode split}, i.e., global and local variable(s) related to modes involved in the task main function;
\item[4-] {\it inter-tasks parameters}, i.e., variable(s) that emphasize the communication between tasks in Fig.~\ref{t:tasks} (e.g., T5-T4 for T5, T2-T6-T1-T3 for T2);  
\item[5-] {\it global params. impact}, i.e., global variable(s) used in performing the respective task functionality;
\item[6-] {\it effect on inter-procedural behavior}, i.e., global impact of function calls; 
\item[7-] {\it effect on the communication} for control navigation and climb tasks (T10 and T12) and {\it arrays} to measure the penalty incurred when transforming array operations into function calls (for T1-T9 and T13), and local impact of specific function calls.
\end{itemize}  
\end{itemize} 

Chisel, when applied on WhileFun programs performs well, partly because of the inter-procedural analysis and partly because the code structure is mode-based. The results for WhileFun are reported in Fig.~\ref{f:bench} (column {\tt red(\%)WhileFun}). On the other hand, we report lower percentages for MIPS code, as shown in Fig.~\ref{f:bench} (column {\tt red(\%)MIPS}) because of several reasons. First, the current version of Chisel does not follow through the memory addresses. Second, any function call in a small sized function involves setting the function stack with registers global and stack pointer, which end-up dominating the code size and yielding longer slices. Third, as reported in~\cite{srinivasan-reps-2016-oopsla}, in general slicing binary code could result in longer slices because of the indirect side-effects via register flags. However, Chisel slicing on MIPS code stays generic and it is work in progress to employ a slicing procedure specialized for low-level languages to produce more precise slices. Using the criteria 6- and 7-, we measure the improvement of an inter-procedural analysis for MIPS that amounts to an additional 20\%-25\% reduction. 

PapaBench is used to evaluate the worst-case execution time (WCET) analysis and to experiment different scheduling models. In these contexts and with respect to the considered benchmark, we used the program slices computed with Chisel for several purposes. For example, we perform the intersection of program slices obtained on criteria such as functionality modes and we identify what are the shared and/or individual behaviors as well as communication patterns at the code level (in particular on WhileFun code). Also, we use the program slices to discover the computationally intensive modes which in turn would impact the task scheduling and the accuracy of the WCET analysis. In this latter case, it is a well-known WCET estimation technique to evaluate the code generated from synchronous designs in two phases: the initialization and the rest.

\section{Concluding remarks and future work\label{sec:conc}}

In this paper we have presented a generic synthesis method for context-updates constructs, from given semantics of programming languages written in Maude. The synthesis strategy follows three stages: the memory policy, the context-updates overapproximations, and the overapproximation refinement. We also integrated our method in Chisel, a Maude tool that can perform generic program slicing. We experimented our extended Chisel with different semantics: WhileFun (imperative) and MIPS (assembly), both of them with different variations, e.g., different memory models and data flow styles. We have also designed test programs to evaluate the efficiency of the produced slices. These experiments correspond to Unmanned Aerial Vehicle applications, which prove that this technique can be applied to real-time programs. Note that for the moment we use these benchmarks in the refinement step.

As ongoing work we focus on a more complex strategy for the refinement step by using more evolved testing strategies. For future work, we plan to extend the language with pointers, hence supporting more complex memory policies based on a more refined memory model. Finally, our aim is to introduce concurrency in the framework so that we can cover and test out proposed methodology on a larger and significant class of programming languages.

{
\bibliographystyle{abbrv}
\bibliography{dak}
}

\end{document}